\documentclass[usenatbib]{mnras}
\usepackage{amsmath}
\DeclareMathOperator\arctanh{arctanh}
\usepackage{graphicx}	
\usepackage{amssymb}	
\title[Perturbative lens reconstruction]{A general method to reconstruct strong gravitational lenses based on the singular perturbative approach.}
\author[Alard, C.]{Alard, C., 
\\
IAP, 98bis Boulevard Arago, Paris \\}
\date{}
\pubyear{2023}
\begin{document}
\label{firstpage}
\pagerange{\pageref{firstpage}--\pageref{lastpage}}
\maketitle
\begin{abstract}
The number of gravitational arcs systems detected is increasing quickly and should even increase at a faster rate in the near future.
This wealth of new gravitational arcs requires the development of a purely automated method to reconstruct the lens and source. A general reconstruction method based on the singular perturbative approach is proposed in this paper. This method generates a lens and source reconstruction directly from the gravitational arc image. The method is fully automated and works in two steps. The first step is to generate a guess solution based on the circular solution in the singular perturbative approach. The second step is to break the sign degeneracy and to refine the solution by using a general source model. The refinement of the solution is conducted step by step to avoid the source-lens degeneracy issue. One important asset of this automated method is that the lens solution is written in universal terms which allows the computation of statistics. Considering the large number of lenses which should be available in the near future this ability to compute un-biased statistics is an important asset. 
\end{abstract}
\section{Introduction}
%
There are basically two main issues with the automated reconstruction of gravitational arcs: (1) what model to choose for a given gravitational arc ? (2) Given  the large parameter space for the models, how to identify some first guess for the solution ? We will show in this paper that the singular perturbative approach offers an efficient solution to these two issues. The paper starts with an introduction to the perturbative singular approach, and continues with a discussion on the proper referential for the reconstruction of the lens. In this referential a first guess for the solution is constructed based on the circular source model in the perturbative approach (see \cite{Alard2007} Eq. (12)). Once this first guess is reconstructed the next step will be to use a general chi-square minimization method to refine the former guess and reach the optimal $\chi^2$ minimum. The degeneracy of the first guess is naturally broken by constructing a refined non circularly symmetric source model. The reconstruction is completed by increasing the potential Fourier order to the required minimal order. The first part of the paper will recall the basics of the singular perturbative approach. This will be followed by a description of the methods and associated equations. The presentation will be illustrated with a number of practical applications by reconstructing simulated gravitational arcs images.
\label{Introduction}

\section{Basics of the singular perturbative approach in gravitational lensing.}
We define the local coordinates ${\bf r}$ and source coordinates ${\bf r_S}$, in a lens potential $\phi$ the lens equation reads,
\begin{equation}
 {\bf r_S} = {\bf r}-\nabla \phi
 \label{Eq_lens_eq}
\end{equation}
The basic idea of the singular perturbative approach is to consider that the un-perturbed situation corresponds to the image of a point at the center of a circularly symmetric potential. Due to its symmetry the image of a point in a circularly symmetric potential is a circle corresponding to the Einstein circle ($R_E$). For convenience we re-normalize the coordinate system to obtain $R_E=1$. As consequence the unperturbed situation is an infinity of points located on the Einstein circle. The choice of this un-perturbed situation allow us to deal with the high non-linearity of strong gravitational lenses in the angular direction. In effect at any angular position $\theta$ an un-perturbed point is present. On the other hand, the perturbation in the radial direction is generally small for gravitational arcs, which allow us to work in the vicinity of the Einstein circle. It is also assumed that strong gravitational arcs are due to small perturbation of a circular potential.Consequently in this framework we have,
\begin{equation}
 \begin{cases}
  \phi(r,\theta)=\phi_0(r)+\epsilon \psi(r,\theta) \\
  r=1+\epsilon dr \\
  {\bf r_S}=\epsilon {\bf r_s}
 \end{cases}
\label{Eq_pot0}
\end{equation}
In Eq. (\ref{Eq_pot0}) $\epsilon$ is a small number. By introducing Eq. (\ref{Eq_pot0}) in the lens equation Eq. (\ref{Eq_lens_eq}) the perturbative lens equation is obtained (see \cite{Alard2007} Eq. (8) and (7)),
\begin{equation}
 \begin{cases}
 \bold {r_s} = (\kappa_2 dr -f_1) \bold u_r - \frac{d f_0}{d \theta} \bold u_{\theta} \\
 f_n(\theta)=\frac{1}{n!} \left[\frac{\partial^n \psi}{\partial r^n} \right]_{r=1} \\
 \kappa_2=1-\left[\frac{d^2 \phi_0}{d r^2}\right]_{r=1}
 \end{cases}
\label{Eq_lens0}
\end{equation}
It is useful to include the source impact parameter ${\bf r_0}$ in the definition of the perturbative fields $f_1$ and $ \frac{d f_0}{d \theta}$ by defining the new fields $\tilde f_1$ and $\frac{d \tilde f_0}{d \theta}$,
\begin{equation}
 \tilde f_i = f_i + {\bf r_0 u_r} ~~~~ {\rm where} ~ i=0,1
\label{Eq_lens2}
\end{equation}
In the continuation of this paper we will assume the impact parameters are included in the field and will not use the ${\tilde f_i}$ symbol to lighten the notation and will use directly the $f_i$ notation.

\subsection{The solution for a circular source in the perturbative approach.}
\label{Sec_round}
An interesting application is to consider the image of a round source contour with radius $R_0$ in the singular perturbative theory.
This model is simple with a direct analytical solution for the images. In effect in this case solving Eq. (\ref{Eq_lens0}) is straightforward and reduces to a second order equation (see \cite{Alard2007}, Eq. 12). 
\begin{equation}
 dr=\frac{1}{\kappa_2} \left[f_1 \pm \sqrt{R_0^2-\left(\frac{d f_0}{d \theta}\right)^2} \right]
\label{Eq_round}
\end{equation}
It is interesting to note that in this circular source model the two perturbative fields have simple physical meanings. The field $f_1$ represents
the mean position of the image for each angular position, while the field $\frac{d f_0}{d \theta}$ is directly related to the image width.
\subsection{Relation to the multipole expansion}
\label{Sec_multi_exp}
A final and interesting point about the singular perturbative expansion is that the perturbative fields are related to the multipole expansion on the Einstein
circle $R_E$ (here re-scale at $R_E=1$).
The multipole expansion of the potential at $R_E=1$ reads:
\begin{equation}
\begin{split}
\psi = - \sum_n \frac{a_n(r)}{r^n} \cos n\theta+\frac{b_n(r)}{r^n} \sin n\theta +c_n(r) r^n \cos n\theta \\
       +d_n(r) r^n \sin n\theta
\end{split}
\label{Eq_pot_multi} 
\end{equation}
The coefficients, $a_n(r),b_n(r),c_n(r),d_n(r)$ of the potential multipole expansion (Eq. \ref{Eq_pot_multi}) are related to the surface
density $\Sigma$ by the following equations,
\begin{equation}
\begin{cases}
a_n=\frac{1}{2 \pi n} \int_0^\pi \int_0^{r=1} \Sigma(u,v) \cos nv \ u^{n+1} dudv \\
b_n=\frac{1}{2 \pi n} \int_0^\pi \int_0^{r=1} \Sigma(u,v) \sin nv \ u^{n+1} dudv \\
c_n=\frac{1}{2 \pi n} \int_0^\pi \int_{r=1}^\infty \Sigma(u,v) \cos nv \ u^{1-n} dudv \\
d_n=\frac{1}{2 \pi n} \int_0^\pi \int_{r=1}^\infty \Sigma(u,v) \sin nv \ u^{1-n} dudv \\
\label{Eq_pot_multi2}
\end {cases}
\end{equation}
This multipole expansion of Eq's (\ref{Eq_pot_multi}) and (\ref{Eq_pot_multi2}) is related to the singular perturbative expansion by Eq. (\ref{Eq_f10_multi}) (see \cite{Alard2009} for the details of the calculations.)
\begin{equation}
\begin{cases}
f_1 =  \sum_n{ n(a_n-c_n) \cos n\theta+n(b_n-d_n) \sin n\theta} \\
\frac{d f_0}{d \theta} =  \sum_n{ -n(b_n+d_n) \cos n\theta+n(a_n+c_n) \sin n\theta}
\end {cases}
\label{Eq_f10_multi}
\end{equation}
It is interesting to note that Eq. (\ref{Eq_f10_multi}) implies a direct relation between the Fourier expansion
of the perturbative fields and the multipole expansion of the potential on the Einstein circle. This also implies that
in this theory the inner and outer contribution to the lens potential can be separated. An example
of a reconstruction with a demonstrative separation of the inner and outer potential contribution can be found in \cite{Alard2010}.
%
%
\section{The advantages of a novel approach based on the singular perturbative method.}
\subsection{introduction}
The singular perturbative approach re-formulate the lens equation Eq. (\ref{Eq_lens_eq}) by using a perturbative expansion of the lens potential in the vicinity of the Einstein circle. The result of this expansion is that the lens equation in this formalism now depends on only 2 fundamental quantities, the perturbative fields. These perturbative fields $f_1(\theta)$ and $\frac{d f_0(\theta)}{d \theta}$ depends only on the angular variable $\theta$. These fields have strong physical meaning and relates directly to the physical features of the gravitational arc (see Sec. \ref{Sec_round}). As a consequence there is no  degeneracy in this representation of the lens. This is not the case for model based representation of the lens. In effect for a given set of perturbative fields there are an infinite number of lens models corresponding to these perturbative fields.
\subsection{Why using the singular perturbative approach to reconstruct gravitational lenses.}
The advantages of this method are obvious, the relation between the image of the source and the perturbative fields is direct. As a consequence constructing a guess solution is straightforward and does not require the costly exploration a large parameter space. Another advantage is the lack of degeneracy of the approach which is again a direct consequence of the correspondence between the image geometry and the perturbative fields. Finally an important asset of this method is its ability to provide a minimal representation of the lens potential (see \cite{Alard2023}). In effect in this approach the order of the perturbative expansion can be limited to the lowest possible order in order to avoid a degeneracy between the source and lens modeling. A detailed study of this source lens degeneracy issue is available in \cite{Alard2023}.
\subsection{Comparison to other methods.}
Conventional methods rely on the adjustment of parametric lens model (see for instance \cite{Brewer} or \cite{Agnello}, \cite{Meneghetti}, \cite{Saha}. This approach has certainly some merit when the model is physically justified (see for instance \cite{Collett}). However for the automated reconstruction of a large set of lens it is difficult to point in the direction of specific models. In this case the complexity of the problem require the construction of analytical lens models with many parameters. The exploration of this large parameter space is a difficult and time consuming task. Furthermore finding a solution in the parameter space does not guaranty that this solution is relevant and is the best solution to the problem. Obviously the analytical modeling of gravitational lenses is prone to degeneracy and the estimation of model parameters may not be independent. A good illustration of this issue is provided in \cite{Etherington} where it is demonstrated that in analytical models with elliptical power law potential and an external shear the estimation of the external shear is un-reliable. This is certainly due to cross dependence between the 2 components of the potential model. Such issues are not present in the singular perturbative approach. In this approach the external and internal components of the potential are easily separated (see Sec. \ref{Sec_multi_exp}). A good illustration
of the reconstruction of the external shear is provided by \cite{Alard2010}. The results of \cite{Alard2010} can be compared to the results obtained by \cite{Tu} using conventional methods. The critical analysis in of the \cite{Tu} results in \cite{Alard2010} indicates many problems with the conventional reconstruction. Some of the associations between the images were wrong and the estimation of the galaxy component in the model was far from the observed galaxy properties. This was not the case in the perturbative model of \cite{Alard2010} where the inner potential inclination matched closely the observed galaxy axis inclination. The problems identified in \cite{Tu} are typical of analytical models with many parameters. The parameter space is very large and finding a minimum  does not guaranty  that an optimal minimum has been found. The problem of cross talk between the parameters presented in \cite{Etherington} is also present here (see also \cite{Saha}). Finally it is interesting to note that in the conventional approach there are no method to prevent the problem of the coupled source lens degeneracy \cite{Alard2023}. This is a serious and concerning issue but it can be tackled easily in the singular perturbative approach. To conclude this paper does not convey the idea that the conventional approach has no merit for the general automated reconstruction of gravitational lenses. When the lens and source system are not too complex like for instance in the case of lensed quasars (see \cite{Schmidt}) the conventional approach may give good results. However even in this case one should be very careful with the results and always investigate to possible degeneracy of the models or the existence of a large family of minima. There are no such issues in the singular perturbative approach and as a consequence it should be the method of choice for general modeling. Furthermore even if conventional methods are used a comparison with current approach should always be a good idea. Finally it is important to note that since the singular perturbative method provide an universal representation of lenses free of degeneracy issues it is a method of choice to analyze lenses from large scale survey. In effect since the perturbative representation of two different lenses is directly comparable free of degeneracies between model components, it is clear that this method will allow us to study the statistical representation of the lens potential for large set of lenses. This would not clearly be the case with the degeneracies and inconsistencies of models with shear (see \cite{Etherington}). This ability to build reliable statistics makes this method unique and of special interest for the large scale surveys.
\subsection{The need for a novel automated implementation of the singular perturbative approach.}
Up to date three reconstruction of gravitational lens systems were made using the singular perturbative approach (see \cite{Alard2009}, \cite{Alard2010} and \cite{Alard2017}). In these reconstructions the guess solution was estimated steps by steps, first by estimating the topology of the solution by observing the image and next by making an appropriate numerical model. None of these steps were automated. The guess solution requires also the estimation of the round solution center, which was done by estimating the position of the center of light for the visible deflector in the image. The Einstein radius was then estimated by considering the closest circle to the images. This value of the Einstein circle was kept constant in the refinement of the guess solution. Here in this new implementation a fully automated method for the estimation of the guess solution will be provided. Furthermore the center of the guess solution will be estimated independently by fitting the center of the closest circle and combined with the conjugate parameters of the $\frac{d f_0}{d \theta}$ field. In this new implementation the round potential center will be refined in the fitting procedure as well as the Einstein radius. To avoid a possible degeneracy the first order parameters of the $f_1$ field will not be refined and rather will be estimated directly by using the $\frac{d f_0}{d \theta}$ first order parameters in the conjugate approach. It is also important to note that in this new implementation the center of gravity of the source is fixed and set at the origin to avoid a possible small source of degeneracy. It eliminates un-necessary parameters and render the reconstruction more efficient. The extensive technical details of this new implementation will be given in the next sections. 
%
\section{The method}
\subsection{Estimating the center of the perturbative reconstruction.}
A basic problem in the perturbative reconstruction of the lens is that the actual center of the unperturbed circular distribution is unknown. In most case the lens itself can be observed and one could use the center of the light distribution in the lens as the center for the reconstruction. However even if this assumption seems reasonable it might be wrong in some cases. It is more interesting to develop a general method without making assumptions and find the center as a parameter of the reconstruction. We will start by considering that we are working in a referential which is slightly shifted form the optimal (true) referential. The effect of the small shift between the two referential on the perturbative fields will be estimated using a perturbative expansion. 
\subsubsection{Effect of a small shift on the perturbative fields.}
Let's define the offset from the optimally centered referential, $\boldsymbol{ \delta_0}$. We will assume that, $|\boldsymbol {\delta_0}| \ll 1$ or equivalently, that  $|\boldsymbol{ \delta_0}|$  is of order
$\epsilon$ with $\epsilon \ll 1$,
\begin{equation}
 \boldsymbol{ \delta_0} = \epsilon \boldsymbol{ \delta} 
\label{Eq_delta}
\end{equation}
The coordinates in the optimal referential and offset referential are respectively $\bf r$ and $\bf u$. The two coordinates are related by,
\begin{equation}
 \boldsymbol{ r = u + \delta_0}
\label{Eq_coord_change}
\end{equation}
The lens equation (Eq. (\ref{Eq_lens_eq})) is transformed using Eq. (\ref{Eq_coord_change}) leading to,
\begin{equation}
\boldsymbol{ r_S=  u+\delta_0} -\nabla \phi\left(\boldsymbol{u+\delta_0}\right)
\label{Eq_lens_eq2}
\end{equation}
By expanding Eq. (\ref{Eq_lens_eq2}) to first order in $\epsilon$ with the help of Eq. (\ref{Eq_delta}) we obtain,
\begin{equation}
\boldsymbol{ r_S = u+\delta_0}-\nabla \left(\phi(\bf{u}) + \boldsymbol{\delta_0} . \nabla \phi(\bf{u}) \right)
\label{Eq_lens_eq3}
\end{equation}
Considering that $\phi=\phi_0+\epsilon \psi$ (see Eq. \ref{Eq_pot0}) and again expanding to first order in $\epsilon$ we obtain,
\begin{equation}
\begin{cases}
\boldsymbol {r_S} = \boldsymbol{u}-\nabla \left(\tilde \phi(\boldsymbol{u})  \right) \\
\tilde \phi(\boldsymbol{u}) = \phi(\boldsymbol{u}) + \boldsymbol{\delta_0}  {\boldsymbol u_r}  \frac{d \phi_0}{ d|\boldsymbol{u}|}-\epsilon \psi_1(\boldsymbol{u})=\phi_0(|\boldsymbol{u}|)+ \epsilon \tilde \psi(\boldsymbol{u})  \\
\tilde \psi(\boldsymbol{u}) = \psi(\boldsymbol{u})+\boldsymbol{\delta} \boldsymbol{u_r}  \frac{d \phi_0}{d|\boldsymbol{u}|}-\psi_1(\boldsymbol{u}) \\
\nabla \psi_1=\boldsymbol{ \delta}
\end{cases}
\label{Eq_lens_eq4}
\end{equation}
According to Eq. (\ref{Eq_lens0}) we have, $f_n=\frac{1}{n!} \frac{\partial \psi}{\partial r}$. The notation $\tilde \psi$ represent the field  $\psi$ once the effect of the offset is taken into account. As a consequence for the offset fields $f_1$ and $\frac{d f_0}{d \theta}$ using Eq. (\ref{Eq_lens_eq4}) we obtain,
\begin{equation}
\begin{cases}
\begin{split}
\tilde f_1 = \left[\frac{\partial \tilde \psi}{\partial u}\right]_{u=1} &= f_1+\boldsymbol{\delta  u_r}\left(-1+ \left[\frac{d^2 \phi_0}{d r^2}\right]_{u=1} \right) \\   &=f_1-\kappa_2 \boldsymbol{\delta u_r}
\end{split}
\\
\frac{d \tilde f_0}{d \theta}=\left[\frac{\partial \tilde \psi}{\partial \theta}\right]_{u=1}=\frac{d f_0}{d \theta}+ 
\boldsymbol{\delta} \frac{\partial \boldsymbol{u_r}}{\partial \theta} \left[\frac{d \phi_0}{dr}\right]_{r=1}-\boldsymbol{\delta}  \boldsymbol{u_{\theta}}=\frac{d f_0}{d \theta}
\end{cases}
\label{Eq_lens_eq5}
\end{equation}
Note that in Eq (\ref{Eq_lens_eq5}) we used $\nabla \phi_1=\boldsymbol{\delta}$ to obtain the radial and angular derivatives of $\phi_1$ (see Eq. (\ref{Eq_lens_eq4})). We also used the fact that $ \left[\frac{d \phi_0}{dr}\right]_{r=1}=1$ which is the consequence of the critical condition
for the unperturbed potential at the scale radius $R_E=1$. Here we used the variable $\kappa_2$ which was already used in the initial presentation of the singular perturbative theory, see \cite{Alard2007}.  It is interesting to note that to the first order  Eq. (\ref{Eq_lens_eq5}) indicates that the field $\frac{d f_0}{d \theta}$ is unaffected by the shift $\boldsymbol \delta$. The field $f_1$ itself is affected by the shift. 
\subsection{The choice of the referential.}
In general the Fourier expansion of the perturbative fields at first order contain four independent terms. Let's define these first order terms as ${\boldsymbol \alpha}$ and ${\boldsymbol \beta}$ for $f_1$ and $\frac{d f_0}{d \theta}$. We have also to include the source impact parameter $\bf r_0$
(see Eq. (\ref{Eq_lens2})). The first order term Fourier of a perturbative fields $F$ is represented by the suffix $F_1$,
\begin{equation}
\begin{cases}
\left[\tilde f_1 \right]_1 = \boldsymbol{(\alpha+r_0) u_r} \\
\left[\frac{d \tilde f_0}{d \theta}\right]_1= \boldsymbol{(\beta+r_0) u_{\theta}}
\end{cases}
\label{Eq_fields_gen}
\end{equation}
We will consider that in our optimal referential these first order ($\boldsymbol{\alpha, \beta}$) Fourier terms are not zero but are the smallest possible. 
If we consider
a slight change in referential by introducing a small shift $\boldsymbol{\lambda}$ by using Eq. (\ref{Eq_lens_eq5}) we obtain,
\begin{equation}
\begin{cases}
\left[\tilde f_1 \right]_1  =  (-\kappa_2 \boldsymbol{\lambda+r_0+\alpha) u_r} \\
\left[\frac{d \tilde f_0}{d \theta}\right]_1 = \boldsymbol{ (\beta+r_0) u_{\theta}}
\end{cases}
\label{Eq_fields_gen2}
\end{equation}
It is interesting to note that Eq. (\ref{Eq_fields_gen2}) implies a fundamental degeneracy in the choice of the referential. In effect one could choose $\lambda$ such that,
\begin{equation}
-\kappa_2 \boldsymbol{\lambda+\alpha=\beta}
\label{Eq_shift}
\end{equation}
Since in the optimal referential model $|\boldsymbol{ \beta}|$ is small then by direct consequence of Eq. (\ref{Eq_shift}) the coefficients at first order for $f_1$ are small two. Thus by this small shift we found another referential where the first order terms are small.
The basic idea behind this choice is that most dark matter halos are not too distant from an isothermal profile. Furthermore it is simple to show that for slightly non-isothermal halos the fields are conjugates for first order Fourier terms.
\subsubsection{Expression of the perturbative fields for slightly non-isothermal potentials.}
\label{Sec_iso_halo}
For a centered circular isothermal potential both perturbative fields are null. When the same potential is slightly off-centered and shifted by ${\boldsymbol \lambda}$ at the first order the potential will be,
\begin{equation}
\phi=|\boldsymbol{r-\lambda}| \simeq |{\boldsymbol r}|+\boldsymbol{ \lambda u_r}
\label{Eq_pot_iso}
\end{equation}
The corresponding perturbative fields can be estimated by using $f_1=\left[\frac{d \phi}{d r}\right]_{r=1}-1+{\bf r_0 u_r}$ and
$\frac{d f_0}{d \theta}=\left[\frac{d \phi}{d \theta}\right]_{r=1}+{\bf r_0 u_{\theta}}$. 
Here $r_0$ is the source impact parameter (see Eq. \ref{Eq_lens2}). 
\begin{equation}
\begin{cases}
\left[\tilde f_1 \right]_1 = \boldsymbol{(\lambda +r_0) u_r} \\
\left[\frac{d \tilde f_0}{d \theta} \right]_1 = \boldsymbol{(\lambda +r_0) u_{\theta}}
\end{cases}
\label{Eq_fields_iso0}
\end{equation}
If we consider a set of slightly of centered isothermal models with corresponding weights potential Eq. (\ref{Eq_fields_iso0}) indicates that the contribution to the perturbative fields will be the weighted sum of of the of the impact parameters of each component. Thus for such slightly perturbed isothermal potentials we see that the perturbative fields will have conjugate expressions and
depends only on two terms. 
 As a consequence we will choose the conjugate field model in the optimal referential. This is a reasonable choice for dark matter halos in this degenerate problem. In any case as long as the shift from the optimal referential is small and of the order of the perturbation size, the specific choice of the referential should not matter and should not affect the final result.  
\subsection{Estimating a first guess for the solution.}
\label{Sec_build_guess}
The process of building the initial guess is de-composed in two steps, first an estimation of the Einstein ring size and center, and second an estimation of the first guess of perturbative fields based on a round source model. 
\subsubsection{Estimating the position and size of the Einstein ring.}
\label{Sec_circ_fit}
The first step is make some estimation of the radius and center of the Einstein circle associated to the lens. The most direct method is to fit a circle to the arc system. To fit a circle to the data we will consider a model where the arc system is a slightly off-centered in the current coordinate system. As a first guess for the center of the circle we may consider the center of gravity the arc system or more directly the center of the image. The actual choice of the guess is not very important as the search for the center and radius of the circle converge quickly even for quite a distant initial guess. For a circle  of radius $R_c$ off-centered by $\bf r_c$ to the first order in $\frac{r_c}{R_c}$ the circle equation reads,
\begin{equation}
r \simeq R_c+{\boldsymbol r_c u_r}=R_c+\left(x_c \cos \theta + y_c \sin \theta \right)
\label{Eq_circle}
\end{equation}
The first step of our processing is to reduce the image of the arc to a set of positions defined by the radius and corresponding angle. The processing starts by choosing a threshold. In the calculations we will consider only the pixels with values above this threshold. In practice the angular domain is divided into a number of bins $N_{\theta}$ and for each bin the mean radius $R_{\theta}$ is estimated. Then the First order Fourier series defined by Eq. (\ref{Eq_circle}) is fitted to the set of positions $(R_{\theta},\theta)$. This fitting procedure estimates the parameters $r_0$, and $\bf r_c$. Note that first we estimate the parameter $\boldsymbol \alpha$ in Eq. (\ref{Eq_circle}) and then we infer $\bf r_c$. Once the estimation of the parameters has been performed the referential is shifted according to the estimation of $\bf r_c$. The the process is iterated, a new estimation is performed and the center is refined. This fitting process is very efficient and converges quickly. The result is a first estimation of the Einstein radius and a first estimation of the lens center.
%
%
The circle fitting procedure converges when the circle is centered in the referential system. When the method converges and the circle is centered in the referential system the first order Fourier terms in Eq. (\ref{Eq_circle}) are zero. 
For a circular source model the field $f_1$ represents the mean position of the images (see Eq. \ref{Eq_round} and Sec. \ref{Sec_round}). As a consequence for a simple round source model the first order Fourier terms of the fields $f_1$ have to be zero in the final referential after the circle fitting procedure has converged.
In a system shifted by a vector $\boldsymbol{\delta}$ from the optimal referential Eq. (\ref{Eq_lens_eq5}) and  Eq. (\ref{Eq_fields_iso0}) implies that,
\begin{equation}
\begin{cases}
\left[\tilde f_1\right]_1 = \left( \boldsymbol{\lambda+r_0} + \kappa_2 \boldsymbol{\delta}\right) \boldsymbol{u_r}=\left(\boldsymbol{\bar r_0} - \kappa_2  \boldsymbol{\delta} \right) \boldsymbol{u_r} \\
\left[\frac{d \tilde f_0}{d \theta}\right]_1= \boldsymbol{(\lambda +r_0) u_{\theta}}=\boldsymbol{\bar r_0 u_{\theta}}
\end{cases}
\label{Eq_fields_iso}
\end{equation}
In Eq. (\ref{Eq_fields_iso}) the degenerate term $\boldsymbol{\lambda + r_0}$ was re-defined by introducing the  $\boldsymbol{ \bar r_0=\lambda + r_0}$. Note that
in the referential defined by the circle fitting procedure the first order terms have to be zero, which by using Eq. (\ref{Eq_fields_iso}) is equivalent to, 
\begin{equation}
 \boldsymbol{\bar r_0}= \kappa_2 \boldsymbol{\delta}
 \label{Eq_ref_lens}
\end{equation}
Our choice for the centering of the arcs is based on an isothermal halo conjugated model (See Sec. \ref{Sec_iso_halo}).For an isothermal halo we have, $\kappa_2=1$, $\bf \bar r_0=r_0$, Eq. (\ref{Eq_ref_lens}) implies that $\delta_0=r_0$ which means that the shift $\delta_0$ corresponds to the source impact parameter. As a consequence for an isothermal potential the circle centering procedure implies that the referential is centered
on the source. As illustrated by the rotation curves the dark matter halos are not far from isothermal thus we the referential will be close to a source centered referential. An interesting point is that at first order the field $\frac{d \tilde f_0}{d \theta}$ is un-affected by the shift and consequently allows a direct
estimation of the pseudo impact parameter $\bf \bar r_0$ (see Eq. (\ref{Eq_ref_lens})).
\subsubsection{Estimating a first guess for the perturbative fields.}
\label{Sec_guess_fields}
A first rough reconstruction of the perturbative fields in this lens centered initial referential will be performed. In the continuation the estimation of the angular bin mean positions and width will be all renormalized by the estimated Einstein radius.  In the circular source model the field $f_1$
is simply the image mean position while the field $\frac{d f_0}{d \theta}$ is related to the image width $\Delta$,
\begin{equation}
\Delta= \sqrt{R_0^2-\frac{d f_0}{d \theta}^2}
\label{Eq_delta}
\end{equation}
The reconstruction of the field $f_1$ is obtained directly by fitting a Fourier series to the mean position $R_{\theta}$ of the binned data at each angular bin position where an image is available. For the field $\frac{d f_0}{d \theta}$ the data to fit will be the image width $\Delta$ (Eq. \ref{Eq_delta}) for each angular bin. An estimation of $\Delta$ will be obtained by counting the number of pixel in each angular bin above the threshold. The number of pixels above the threshold is a direct estimation of the bin surface from which the effective width $\Delta$ can be easily estimated. In practice the number of pixel above a threshold is an integer quantity and as a consequence may not be accurate for a small number of pixels. To improve the accuracy of this estimate the number of pixels just below the threshold and just above the threshold are estimated (see Eq. (\ref{Eq_pot_nfw})). An interpolation of these two nearby pixel counts provide an accurate estimation of the pixels number at the threshold for each angular bin $C_{\theta}$. 
\begin{equation}
\Delta= \frac{N_{\theta} C_{\theta}}{2 \pi R_c^2 }
\label{Eq_delta2}
\end{equation}
It is in principle possible to evaluate the field $\frac{d f_0}{d \theta}$ by relating the numerical data of Eq. (\ref{Eq_delta2}) with equation  Eq. (\ref{Eq_delta}). However in practice a direct evaluation of $\frac{d f_0}{d \theta}$ using Eq's (\ref{Eq_delta}) and (\ref{Eq_delta2}) is problematic for two reasons. The first reason is that due to the square in Eq. (\ref{Eq_delta}) the sign of $\frac{d f_0}{d \theta}$ is unknown. Each time the function $\frac{d f_0}{d \theta}$ cross the zero line the sign may or may not change. The field $\frac{d f_0}{d \theta}$ is zero near the center of each image, thus if there are many images
the uncertainty on the sign means that the number of potential solution to explore is large. And furthermore the second issue is that the management of the noise is very delicate due to the square root in Eq. (\ref{Eq_delta}). An optimal estimation of $\frac{d f_0}{d \theta}$ requires a fitting procedure based on a least-square model. The Fourier series are the simplest and most direct non parametric reconstruction of the perturbative fields since they are directly related to the multipole
expansion on the Einstein circle (see Eq. (\ref{Eq_pot_multi})). To construct the initial guess it is important to reduce the complexity of the solution to a minimum and thus to consider an expansion in Fourier series of the fields of lower order. A typical number for the order of the Fourier $N_F$ expansion is $N_F=2$ or $N_F=3$ depending on the complexity of the solution. It would be very unusual and quite impractical to reach $N_F=4$ for the initial guess. Another point is the existence of a possible mass-sheet degeneracy related to the scaling coefficient $\kappa_2$ (see Eq. (\ref{Eq_lens0})). In order to cancel the effect of the re-scaling coefficient $\kappa_2$ it is useful to introduce the correlation coefficient which is scale free rather than a least-square estimate. As a consequence the estimation of $\frac{d f_0}{d \theta}$ will be performed by maximizing the correlation coefficient between the two expression of $\Delta$, the analytical expression in Eq. (\ref{Eq_delta}) and the numerical data of Eq. (\ref{Eq_delta2}). The order of the Fourier series is low and the cost of computing the correlation coefficient for the array of binned data is low since in practice the number of bins is less than one hundred. As a consequence  it is definitely practical to explore the parameter space to estimate a model for $\frac{d f_0}{d \theta}$. The field $\frac{d f_0}{d \theta}$ has no zero order term since it is a derivative, thus at order 2 the number of coefficients is 4 and 6 at order 3. The parameter space is explored by generating a series of random numbers in a typical range. The range to investigate is by nature of a fraction of unity. In practice the range at order 1 is about 0.5, and decrease by a scale factor corresponding to the order
at higher order. The assumption of decreasing order is justified by the fact that the power of the Fourier series at some higher order will be close to zero. Thus
between the first order and this higher order a decrease is observed and we make the choice of the simplest model, a linear decrease. In practice this choice is notation very important since we start with a very broad range at order 1 which is much larger than practical cases. This model of linear decrease can thus be considered as an upper bound. At order 2 the exploration of the parameter space is performed using $10^6$ set of random number. The corresponding computing time is only of about a few seconds on a basic PC for 75 angular bins. Considering the low computing cost the number of random sets can be easily increased if higher order Fourier solutions have to be explored. 
\subsubsection{Illustration using a numerical simulation.}
\begin{figure*}
\includegraphics[width=14cm]{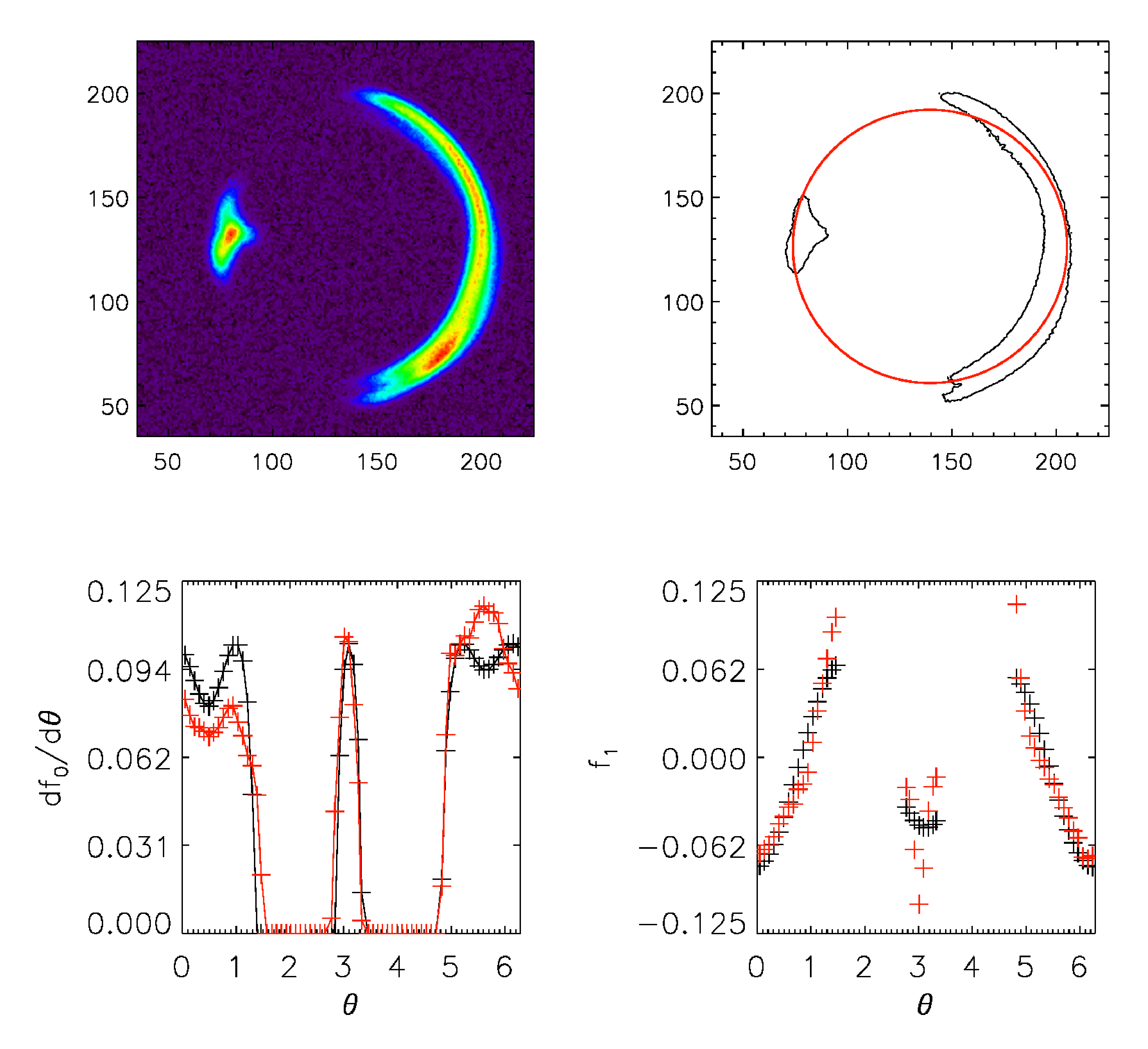}
\caption{Upper left panel: the simulated image corresponding to simulation $A_1$. Upper right panel: the reconstructed image. Lower left panel: the first guess estimation for the field $\frac{d f_0}{d \theta}$. lower right panel: the first guess estimation for the field $f_1$.}
\label{Fig_1}
\end{figure*}

The application of this method to a numerical simulation of a cusp configuration for an elliptical NFW potential is presented in Fig. (\ref{Fig_1}) (upper left panel). The details of this numerical simulation are available in Sec. (\ref{Sec_simu}).
The different steps of the reconstruction are illustrated in the upper right panel
 with the circle fitting procedure, and in the lower panel with the first guess for the perturbative fields based on the round source model.
\subsection{Refining the guess and converging to an optimal solution.}
Once a a first approximate solution based on the circular source model has been obtained for the fields the problem is to refine this solution and find the optimal solution. In practice the approximate solution will be used as a starting guess for a full minimization in the multi-parameter space. However the first thing to consider in the search for an optimal solution is to define an optimal referential. An optimal referential should minimize the non-circular parts of the lens potential and as a consequence should be closely centered on the lens. However the circle fitting procedure implies that the first guess is defined in a source centered referential (see Sec. (\ref{Sec_circ_fit}) and Eq's (\ref{Eq_fields_iso}) and (\ref{Eq_ref_lens})). Consequently the first step should be to move from the source centered referential where we defined the initial guess to a lens centered referential. This task is facilitated by the fact that the field $\frac{d f_0}{d \theta}$ is unchanged by a small shift (see Eq. (\ref{Eq_lens_eq5})). When expressed in a referential shifted by
$\bf \delta_0$ the first order term of $\frac{d f_0}{d \theta}$ are precisely the pseudo-impact parameter $\bar \bf r_0$
(see Eq. (\ref{Eq_fields_iso})). Thus from the guess solution for $\frac{d f_0}{d \theta}$ we can estimate
the pseudo-impact parameter $\bar \bf r_0$. A problem is that the guess solution is estimated from the circular solution of Eq. (\ref{Eq_round}). In this equation the square of $\frac{d f_0}{d \theta}$ appear which mean that a degeneracy
on sign of $\frac{d f_0}{d \theta}$ is present. This means that the sign of $ \bf \bar r_0$ is unknown as well as the higher order Fourier terms of the field $\frac{d f_0}{d \theta}$. In practice it mean that the two solutions corresponding
to the two signs of $\frac{d f_0}{d \theta}$ have to be explored. The circular source solution is insensitive to the sign of $\frac{d f_0}{d \theta}$ but this is not the case for a non-circular source. As a consequence the $\chi^2$ will have to be evaluated for each sign of $\frac{d f_0}{d \theta}$ and the minimal $\chi^2$ will point point to the proper sign and break the degeneracy.
\subsubsection{Reconstructing the lens and source by minimizing the $\chi^2$.}
\label{Sec_refine_guess}
\subsubsection{The parameters of the lens reconstruction.}
\label{subsec_params}
The parameters that will have to be refined are the following, the Einstein radius $R_C$, the center of the unperturbed circular potential $\bf r_c$, the pseudo impact parameter $\bar \bf r_0$, and the higher order Fourier terms ($N>1$) of the perturbative fields. Note that the first order Fourier terms of the perturbative fields are conjugated (see Eq.  (\ref{Eq_fields_iso}) here for $\bf \delta_0=0$), and as a consequence there are only two independent terms. The first order terms for $f_1$ are directly inferred from the first order terms of $\frac{d f_0}{d \theta}$ by conjugation.
\subsubsection{The source reconstruction procedure.}
\label{subsec_source}
The source is represented by the linear combination of 2D Gaussian weighted polynomials (see \cite{AlardLupton}). The 2D amplitude
in the source plane represented by the coordinates $(x_s,y_S)$ is,
\begin{equation}
I_S=\sum_{i=0,N} \sum_{j=0,N-i} a_{i,j} \exp(-\left(\frac{x_S^2+y_S^2}{r_0^2} \right) x^i y^j
\label{Eq_source}
\end{equation}
The modelisation of complex sources which are frequently encountered in gravitational lenses requires a higher order for the expansion in Eq. (\ref{Eq_source}). In practice to avoid the source-lens degeneracy the initial source expansion should be of low order, but should be increased to reach a final higher order with typically $N=8$ or $N=10$. An initial estimation of $r_0$ is obtained by estimating the residual to the fit of a circle to the arc (see Sec \ref{Sec_circ_fit}).
Once a first fit of the lens has been performed $r_0$ is evaluated directly by computing the size of the actual source solution. This process of refining $r_0$ is iterated for each steps of the fitting procedure.
\subsubsection{Fitting the image and estimating the $\chi^2$.}
\label{subsec_chi2}
For a model of the perturbative fields to a given Fourier order the use of the lens equation allows (Eq. (\ref{Eq_lens0})) allows to transport each component of the source model (Eq. \ref{Eq_source}) to the lens plane. Each corresponding image
of the source component is then convolved with a Gaussian PSF model. The set of convolved images is then fitted to the arcs system images using the linear least-square method (see \cite{Warren_Dye}). This fitting procedure allows the evaluation of the $\chi^2$ corresponding to the specific model for the perturbative fields.
\subsection{Iterating the procedure.}
\label{Sec_iter}
Starting from the initial circular source guess we will make a search for the minimal $\chi^2$ solution. The minimization of the $\chi^2$ in the parameter space of the lens (see Sec. (\ref{subsec_params}) ) is performed using the simplex method \cite{Nelder}.For each set of parameters the $\chi^2$ is evaluated according to the procedure described in Sec. (\ref{subsec_chi2}). Due to the degeneracy on the sign of $\frac{d f_0}{d \theta}$ the search for the minimal $\chi^2$ is conducted for both negative and positive signs. The choice between the 2 solutions with different signs is made by selecting the solution with minimal $\chi^2$.
\subsubsection{Avoiding the source lens degeneracy problem.}
\label{Sec_deg}
The source lens degeneracy may be present due to the interaction and degeneracy between higher order source and lens terms. As shown in \cite{Alard2023} a solution to this issue is to consider the minimal solutions. These minimal solutions are among the class minimal $\chi^2$ solutions those with the lowest possible order in the Fourier expansion of the fields. It is demonstrated in \cite{Alard2023} that these minimal solutions can be constructed by starting the search for the solution to the lowest possible order. A typical number to reconstruct the guess solution is $N=2$ or $N=3$ for the Fourier expansion of the fields. This will be also the order at which the solution is reconstructed for the first step
of the refinement. Once the sign degeneracy is broken the order of the solution is increased progressively to reach the highest order for the solution (typically $N=4$ or $N=5$). Note that in similarity to the reconstruction process of the guess solution the size of the simplex corresponding to the higher order terms is of decreasing amplitude. Here the same rule applies, with a decrease in the amplitude proportional to order. It is possible to relax this constraint for some specific case, however it is better to try using this modulation of the higher order terms first. 
\subsection{Application of the method to a set of numerical simulations.}
In this paragraph we will demonstrate the refinement of the guess solution using a set of 4 numerical simulations. 
\subsubsection{Description of the numerical simulation.}
\label{Sec_simu}
The four numerical simulations corresponds to a cusp like configuration for 4 different potential types. The different potential models are a second order
lens potential (elliptical) a fourth order lens potential (quadratic) and an external shear. The combination of the elliptical and quadratic potential with the external shear give the four potential models (see Table \ref{Tab_1}) and the four corresponding arcs simulations $A_{i=1,n}$.
\begin{table}
\begin{tabular}{|c | c | c |}
\hline
 Arcs Simulation  & Potential Fourier order  & External Shear   \\
 \hline
 $A_1$  & 2 & no    \\
 $A_2$  & 2 & yes  \\
 $A_3$  & 4 & no  \\
 $A_4$  & 4 & yes \\
\hline
\end{tabular}
\label{Tab_1}
\caption{The four types of simulations.}
\end{table}
\subsubsection{The potential models in the simulations}
The potential models are based on the NFW potential described in Eq. (28) in \cite{Alard2007}, and \cite{Bartelmann1996}. The iso-contour of this potential are slightly distorted from circularity. The distortion is expressed using Fourier series to order $N$. The coefficients of the Fourier series representing the distortion are
represented by $\eta_i$. The lowest order corresponds to an elliptical distortion, $N=2$ (2 coefficients), and the highest order in this simulation corresponds to $N=4$ (quadratic).
\begin{equation}
\begin{cases}
\phi \propto \frac{1}{2} \log^2 \left( \frac{x}{2} \right) -2 \arctanh^2 \left( \sqrt{\frac{1-x}{1+x}} \right) \\
x=r \left(1+\sum_{k} \eta_{[2(k-2)]} \cos(k \theta) + \eta_{[2(k-2)+1]} \sin(k \theta) \right)
\end{cases}
\label{Eq_pot_nfw}
\end{equation}
The shear is external and corresponds to the effect of mass situated outside the Einstein circle. The shear is related to order 2 terms in the potential and as a consequence to order 2 Fourier series terms in the potential expansion. Using Eq. (\ref{Eq_f10_multi}) for an external shear of order 2 he generic expression for the fields is,
\begin{equation}
\begin{cases}
\left[f_1 \right]_{shear}=S_0 \cos 2 \theta -S_1 \sin 2 \theta \\
\left[\frac{d f_0}{d \theta} 
\right]_{shear}=-S_1 \cos 2 \theta -S_0 \sin 2 \theta 
\label{Eq_shear}
\end{cases}
\end{equation}
For the two numerical simulation with shear we have $S_0=S_1=0.05$.
\subsubsection{Building the simulations.}
For each of the potential models $A_i$ the image of the source is reconstructed using ray-tracing. 
The source in this simulation is complex and quite distant from a round source (see Fig. (\ref{Fig_f2_no}) lower panel for an image of the source contours). The image of the source was convolved
with a Gaussian PSF with a FWMH of 2.5 pixels. The final image was produced by adding Poisson noise to the convolved image. 
\subsubsection{Elliptical second order potential with no shear.}
\begin{figure*}
\includegraphics[width=14cm]{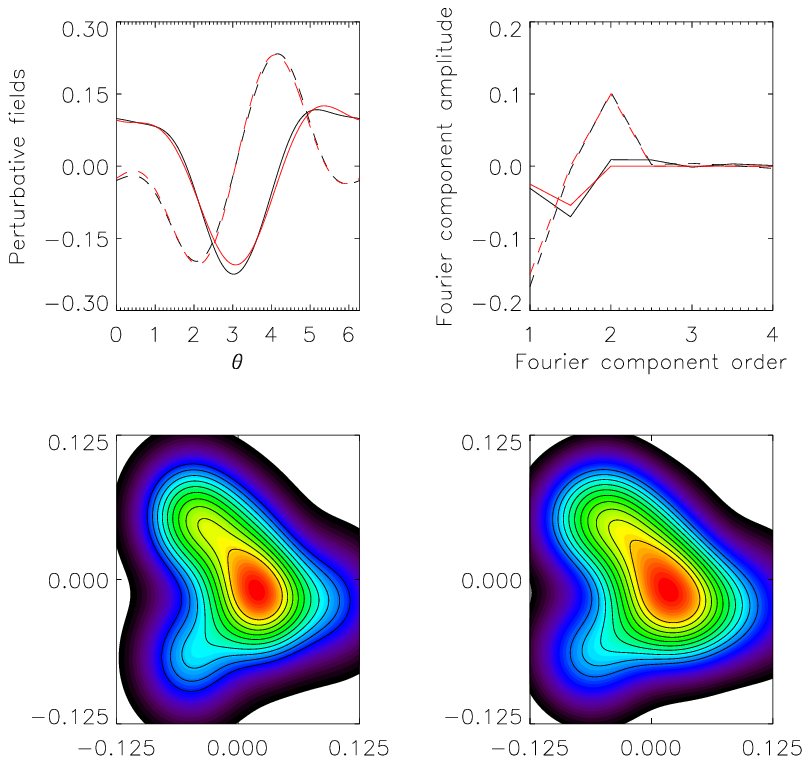}
\caption{Upper left panel: a comparison between the reconstructed fields (black line) and the true solution (red line)  Upper right panel: The amplitude of the Fourier series expansion of the fields as a function of the order. In both figures the field  $\frac{d f_0}{d \theta}$ is represented as a dotted line,  while the $f_1$ is represented by a continuous line. The lower left panel represents the true source while the lower right panels represents the reconstructed source.}
\label{Fig_f2_no}
\end{figure*}

\begin{figure*}
\includegraphics[width=14cm]{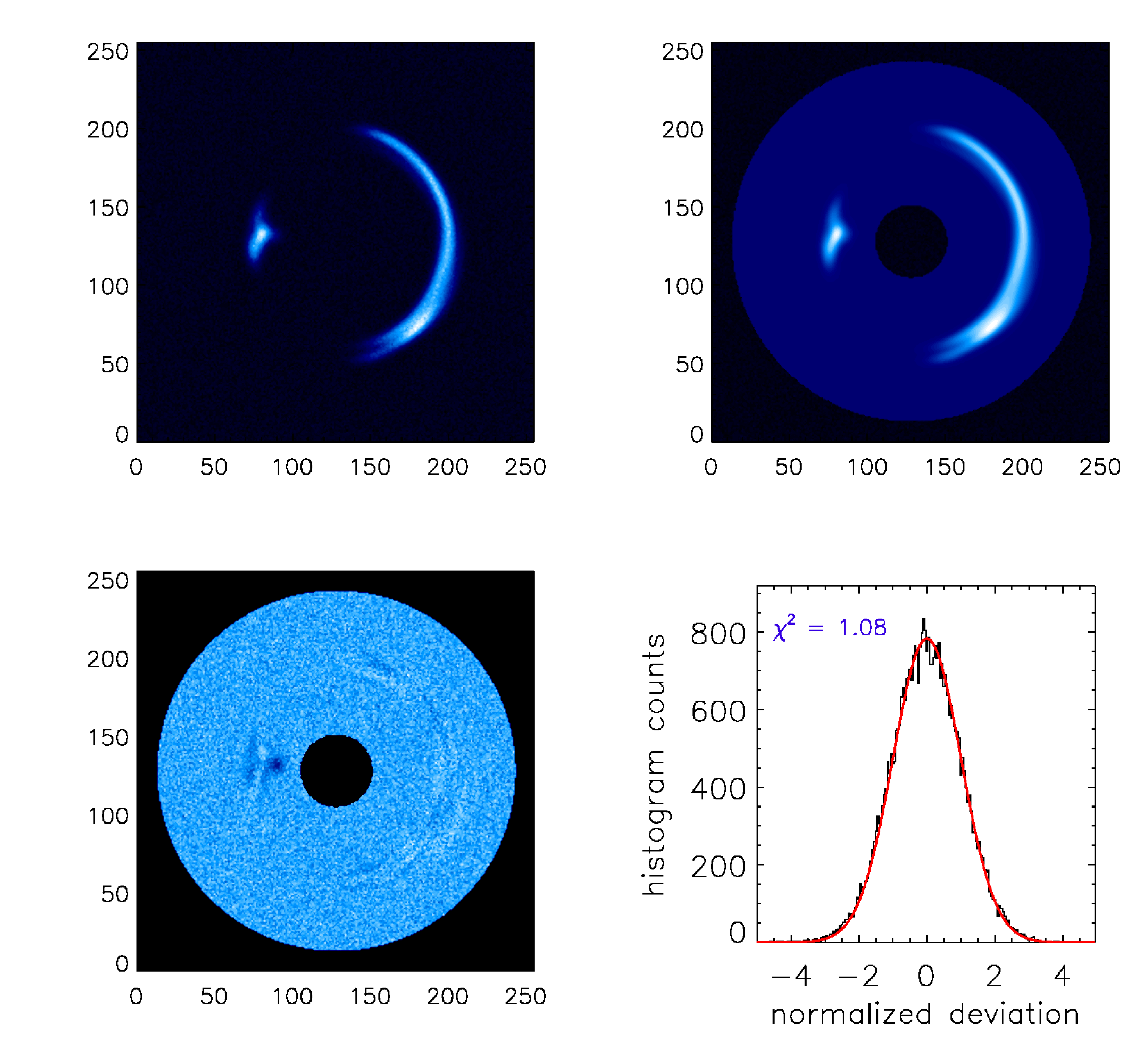}
\caption{Upper left panel: the image of the arc system corresponding to simulation $A_1$. Upper right panel: the reconstructed image obtained after the final iteration. Lower left panel: subtraction between the true and reconstructed image. Lower right panel: an histogram the noise normalized by the theoretical noise expectation. Note that the histogram (red line) is quite close to the Gaussian expectation (black line). }
\label{Fig_i2_no}
\end{figure*}

This simulation corresponds to case $A_1$ in Table (\ref{Tab_1}). The process of building this guess solution was described in Sec. (\ref{Sec_build_guess}). The resulting guess solution for this simulation is illustrated in Fig. (\ref{Fig_1}). Starting from this guess solution the solution is refined in a series of steps with each time an increase
of the order for the Fourier series expansion of the potential. The refinement operates according to the procedure described in Sec. (\ref{Sec_refine_guess}). The refinement process starts by exploring the two solutions corresponding to each sign of the field $\frac{d f_0}{d \theta}$ (see Sec (\ref{Sec_iter})). In this first iteration the Fourier order of the potential is 2. From the two solutions the solution with minimal $\chi^2$ is retained. This solution is refined by increasing the Fourier order of the potential to 3 and subsequently to 4 to avoid the source degeneracy problem (see Sec. (\ref{Sec_deg})). The refinement process is stopped at order $N=4$. Not that for minimality reason the reconstruction could have been stopped at order $N=2$ in this case. However in order to provide a comparison of the different lens reconstructions in a self consistent scheme all lens are reconstructed to order 4. The reconstruction of the lens to a higher is also interesting since it demonstrates the stability of the solution at higher order.
All solutions were built using an eight order Gaussian-polynomial solution for the source (See Sec. (\ref{Sec_source})).The distortion vector corresponding to simulation $A_1$ is,
\begin{equation}
\eta=[-0.05,0,0,0,0,0]
\end{equation}
The final result is presented in Fig.'s (\ref{Fig_f2_no}) and (\ref{Fig_i2_no}).
\subsubsection{Elliptical second order potential with shear.}
\begin{figure*}
\includegraphics[width=14cm]{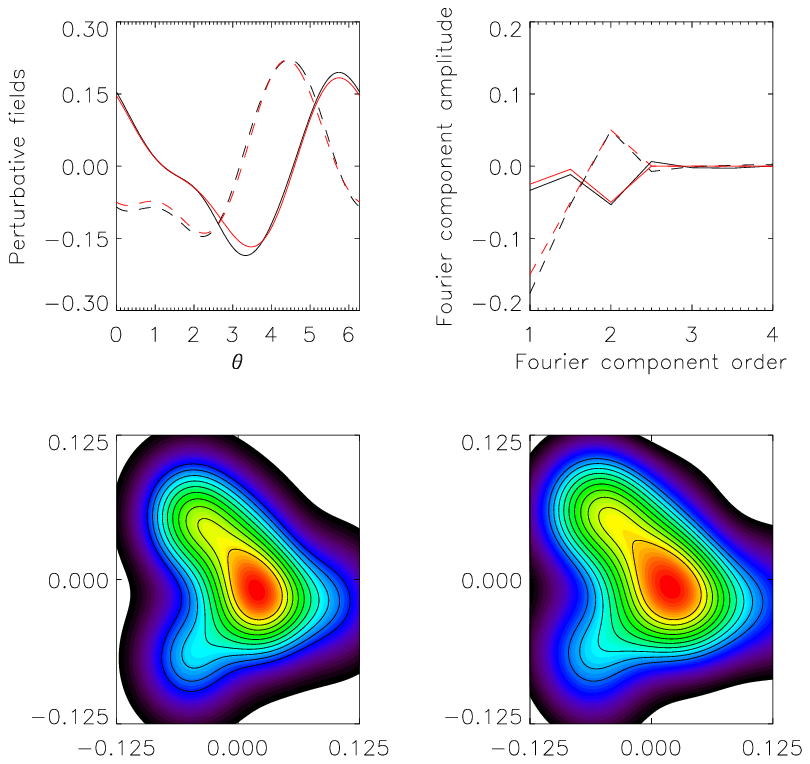}
\caption{Same as a Fig. (\ref{Fig_f2_no}) but this time for simulation $A_2$.}
\label{Fig_f2_w}
\end{figure*}

\begin{figure*}
\includegraphics[width=14cm]{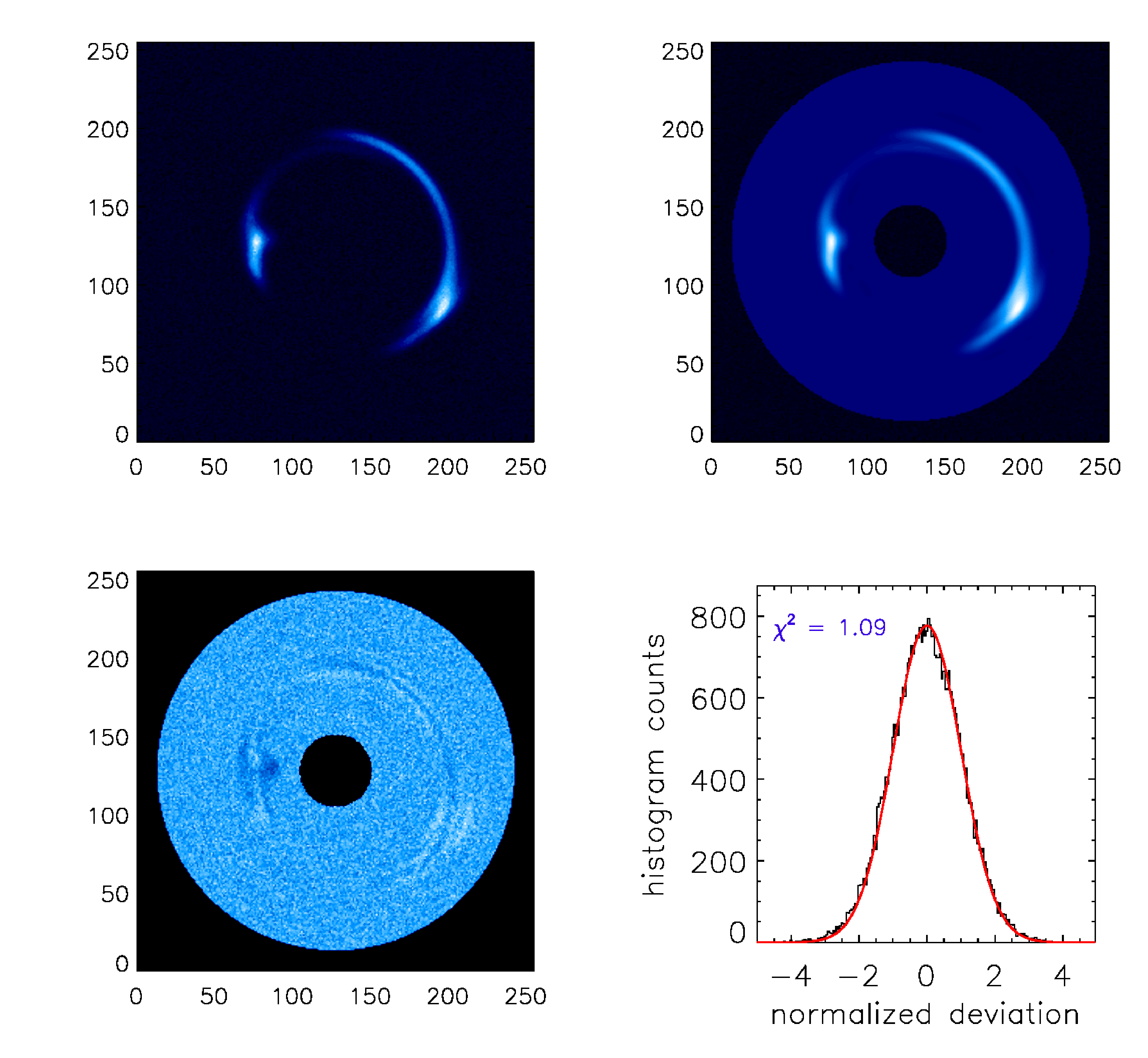}
\caption{Same as a Fig. (\ref{Fig_i2_no}) but this time for simulation $A_2$.}
\label{Fig_i2_w}
\end{figure*}

The simulation $A_2$ is similar to simulation $A_1$ except for the presence of an external shear (see Eq. (\ref{Eq_shear})). The construction of the guess solution and of the final refined solution follows the same process that was already described for simulation $A_1$. 
The final result of the refinement process is presented in Fig.'s (\ref{Fig_f2_w}) and (\ref{Fig_i2_w}).
\subsubsection{Fourth order potential with no shear.}
\begin{figure*}
\includegraphics[width=14cm]{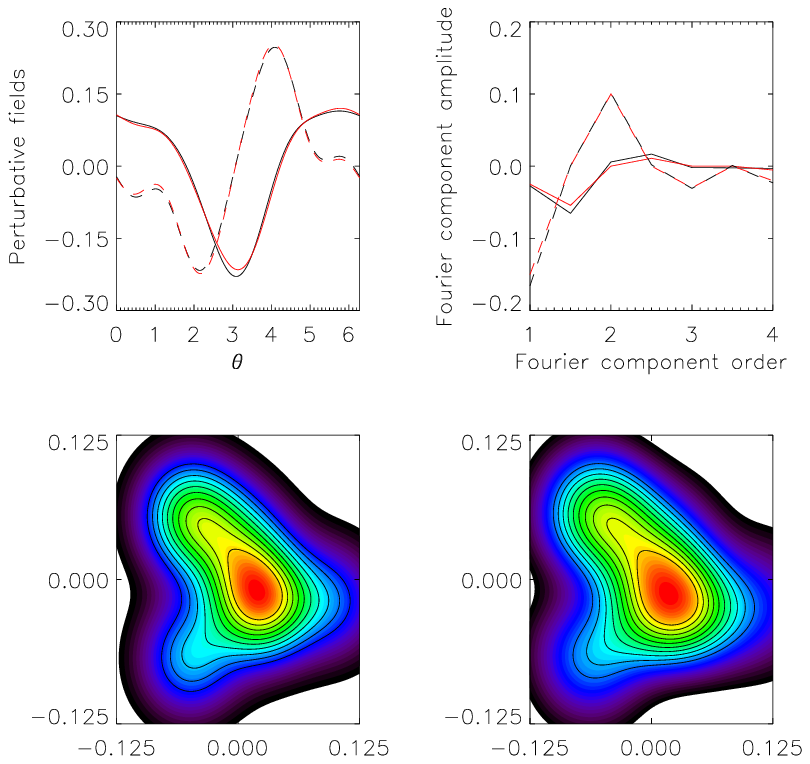}
\caption{Same as a Fig. (\ref{Fig_f2_no}) but this time for simulation $A_3$.}
\label{Fig_f4_no}
\end{figure*}

\begin{figure*}
\includegraphics[width=14cm]{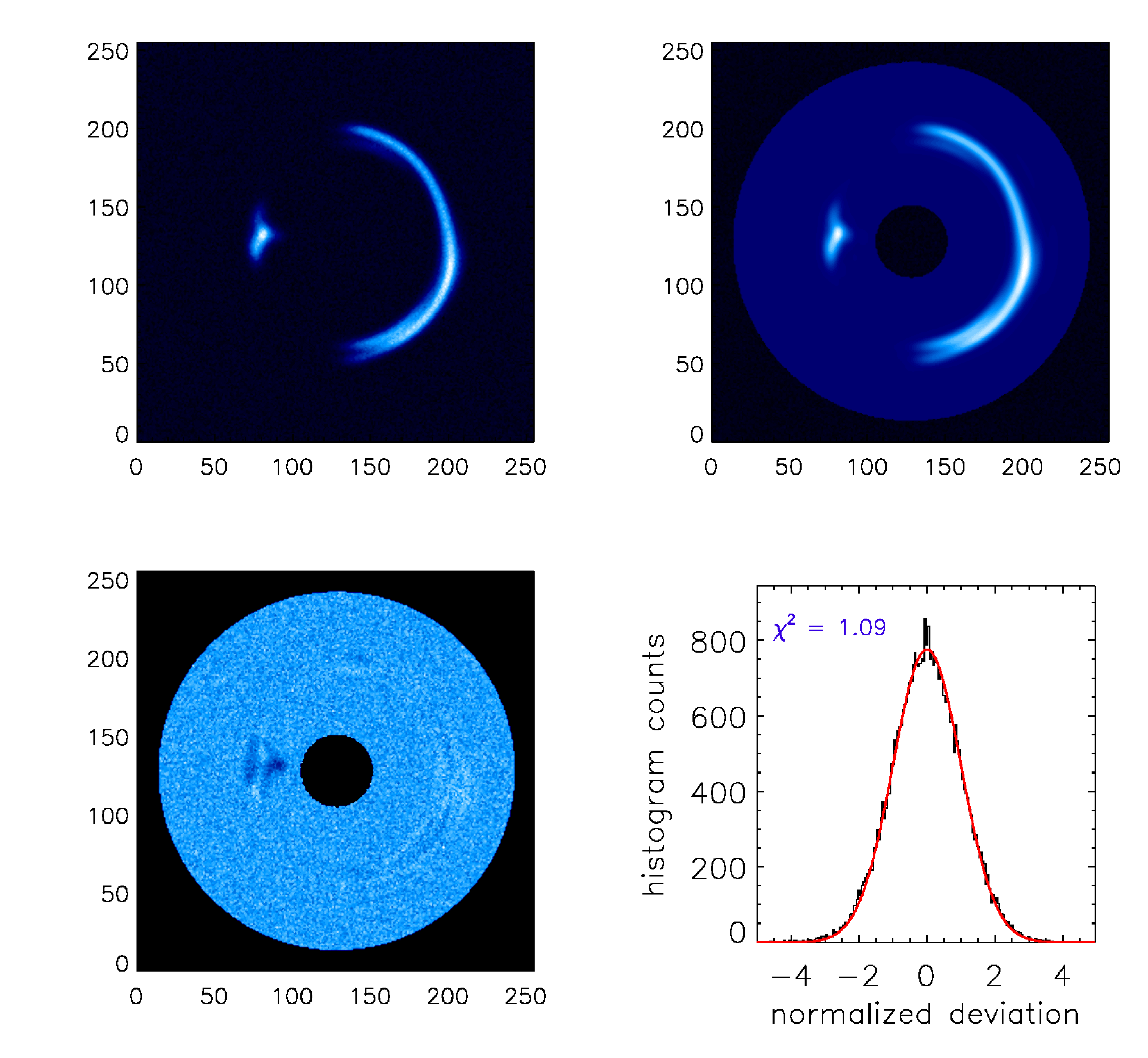}
\caption{Same as a Fig. (\ref{Fig_i2_no}) but this time for simulation $A_2$.}
\label{Fig_i4_no}
\end{figure*}

Here simulation $A_3$ corresponds to an NFW potential with a fourth order Fourier distortion of the iso-contours . The corresponding distortion vector $\eta_i$   is defined in Eq. (\ref{Eq_pot_nfw}). Here again the construction of the guess solution and of the final refined solution follows the same process that was already described for simulation $A_1$. For simulation $A_3$ we have,
\begin{equation}
\eta=[-0.05,0,0.01,0,0,-0.005]
\end{equation}
The final result of the refinement process is presented in Fig.'s (\ref{Fig_f4_no}) and (\ref{Fig_i4_no}).
\subsubsection{Fourth order potential with  shear.}
\begin{figure*}
\includegraphics[width=14cm]{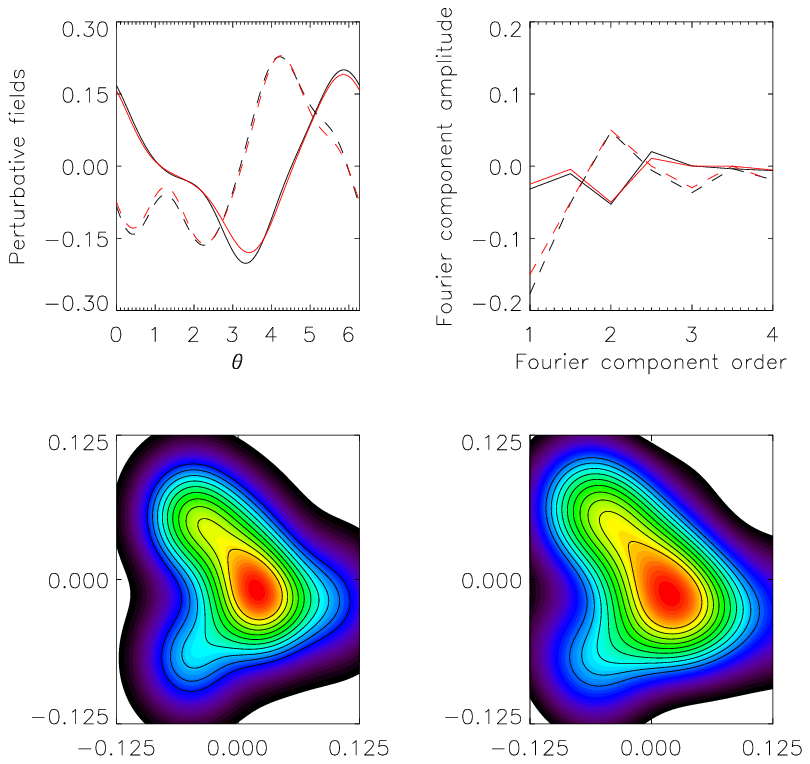}
\caption{Same as a Fig. (\ref{Fig_f2_no}) but this time for simulation $A_4$.}
\label{Fig_f4_w}
\end{figure*}

\begin{figure*}
\includegraphics[width=14cm]{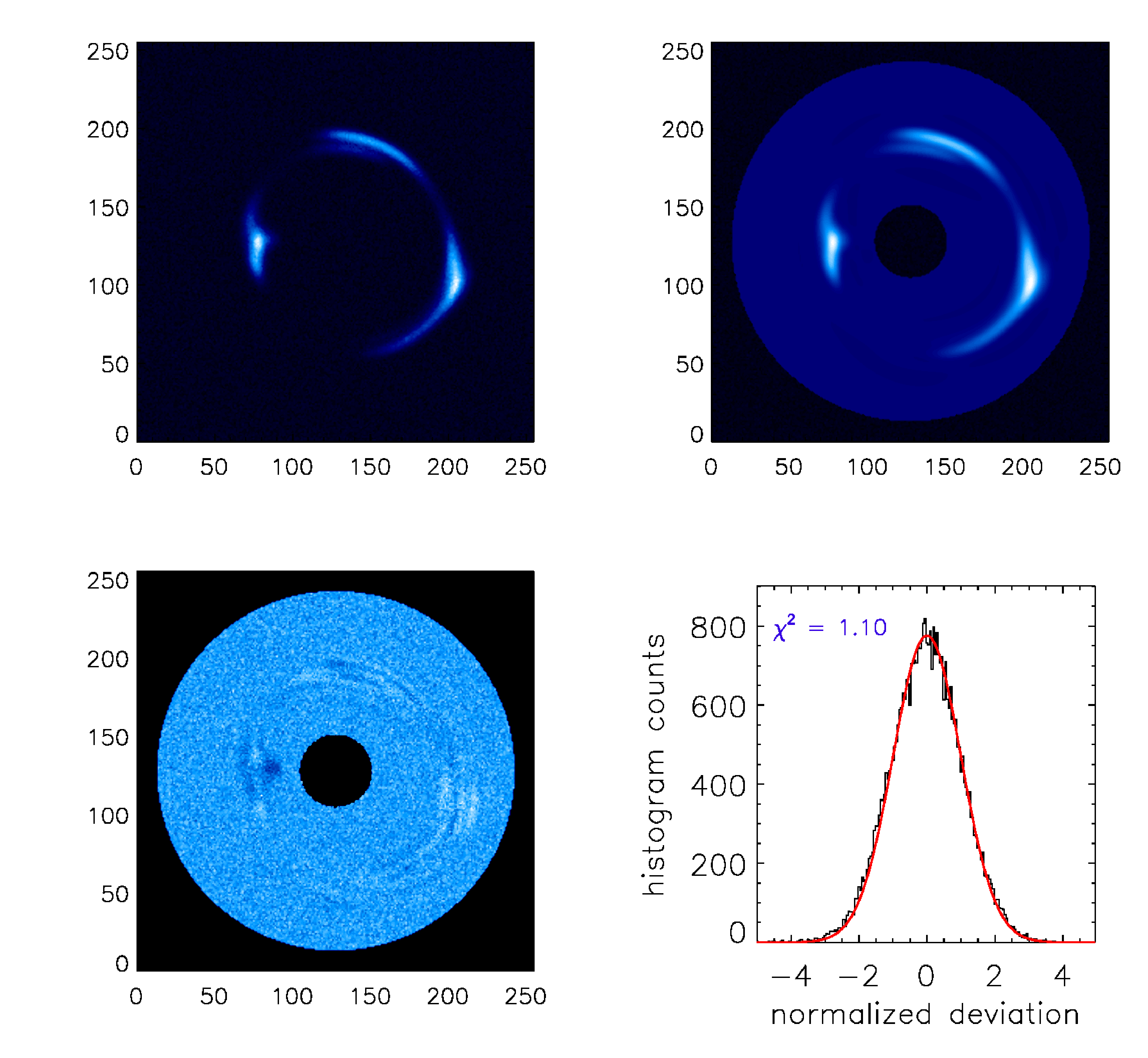}
\caption{Same as a Fig. (\ref{Fig_i2_no}) but this time for simulation $A_4$.}
\label{Fig_i4_w}
\end{figure*}

Simulation $A_4$ corresponds to simulation $A_3$ with the addition of an external shear(see Eq. (\ref{Eq_shear})). Here again the construction of the guess solution and of the final refined solution follows the same process that was already described for simulation $A_1$. The final result of the refinement process is presented in Fig.'s (\ref{Fig_f4_w}) and (\ref{Fig_i4_w}).
\subsubsection{Result of the simulations.}
It is clear form the results of the simulations see Fig's (\ref{Fig_i2_no}), (\ref{Fig_i2_w}), (\ref{Fig_i4_no}) and (\ref{Fig_i4_w}) that all configurations and even the more complex were properly reconstructed using the automated method presented in this paper. No assumptions were made about the arcs themselves and the procedure produced the lens and source
modeling directly from the images. All results were produced using the same universal singular perturbative approach which  suggests that an automated analysis of a large data set will now become possible with this method. Interestingly the universality of the modeling will also render possible a statistical analysis of large data sets.
%
%
\section{Application of the method to the Horseshoe lens SDSS J1148+1930.}
\label{Sec_horsehsoe}
\subsection{Pre-processing of the data.}
\begin{figure*}
\includegraphics[width=16cm]{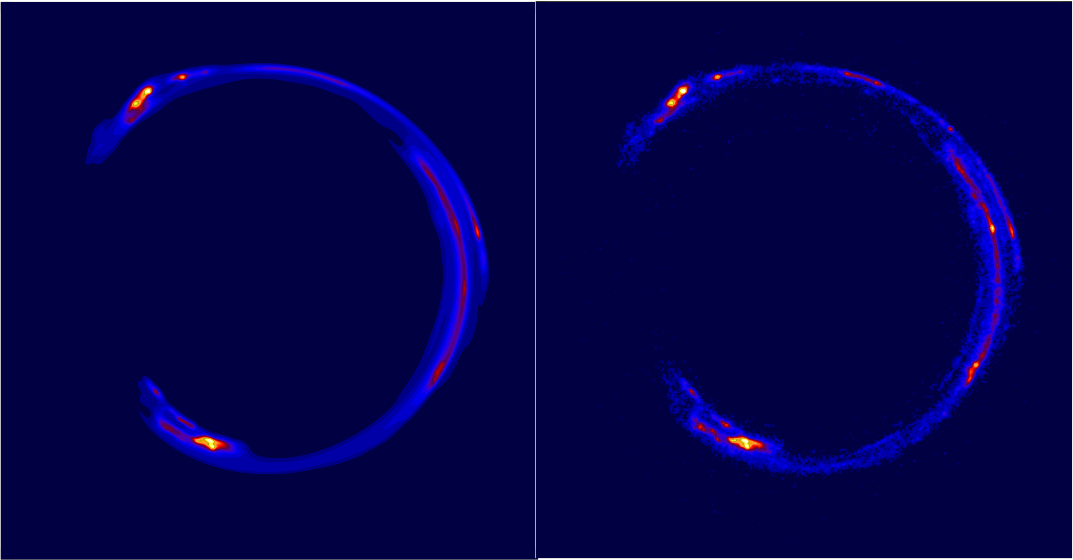}
\caption{The HST image of the horseshoe lens in the f606w photometric band (right side). On the left side next to the actual HST image of the lens is the final reconstruction of the lens using the accurate high resolution source model (see Sec. \ref{Sec_source}).}
\label{Fig_horseshoe_images}
\end{figure*}

The reconstruction of the arc will be performed on the HST image of the Horseshoe lens taken with the  f606w filter.
In order to isolate the arc and cancel any additional contribution from foreground sources a circular area around the arc was selected.
Pixels out of the selected circular zone were set to zero (see Fig. (\ref{Fig_horseshoe_images})).
As already noted by \cite{Alard2017} the subtraction of the contribution from the wings of the central galaxy is very weak and already at the noise level in the f814w filter. In the  f606w photometric band the contribution of the wings is even weaker thus trying to fit a sersic profile to the central galaxy light profile may not be the best option. In effect the extrapolation of the galaxy light profile using an analytical model involves some uncertainties. These uncertainties are at least of the order of the correction. As a consequence, rather than introducing some unknown systematic bias the contribution of the galaxy wings was ignored.
\subsection{Estimation of the first guess}
\begin{figure*}
\includegraphics[width=14cm]{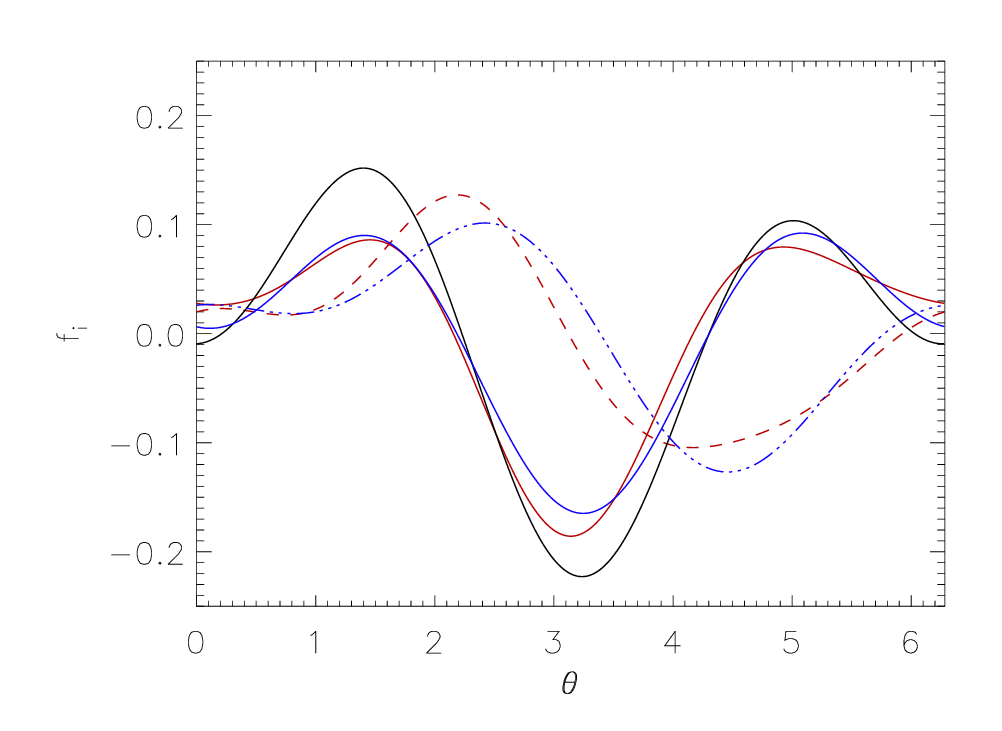}
\caption{The solutions for the perturbative fields. The field $f_1$ is represented by dotted lines (blue color order 2, red color order 3). The field $\frac{d f_0}{d \theta}$ is represented by solid lines (black color first guess, blue color order 2, red color order 3).}
\label{Fig_horseshoe_field1}
\end{figure*}

The process of estimation of the first guess solution is described in detail in Sec. (\ref{Sec_build_guess}). The first step in this process is the estimation of the best fitting circle. Since the lens image cover almost the full angular domain  this task is straightforward. This fitting procedure allows a first estimation of the value of the Einstein radius and the approximate estimation of a source centered referential.The next step is the estimation of the arc mean radial position and width. The mean radial position of the arc in bins of angular positions is estimated (in the referential defined by the circle fitting procedure). The width of the arc is also estimated for each angular bin. The total number of angular bin used was 70. The estimation of the perturbative fields is performed using the circular source approximation. The field $f_1$ is estimated directly by fitting a Fourier series to the mean radial position of the arc. The field $\frac{d f_0}{d \theta}$ is evaluated by fitting a Fourier series to the estimated width of the arc. Since the equation relating the image width to $\frac{d f_0}{d \theta}$ is slightly non-linear (see Eq. (\ref{Eq_round})) the parameter space of the Fourier series is fully explored using monte-carlo generation of the parameters. Since the cost of computing the width from an estimation of $\frac{d f_0}{d \theta}$ is very low the computing time required to explore the full parameter space is small (less than a minute on a basic laptop computer). The Fourier series expansion for the first guess was limited at order two. The first guess for the field $\frac{d f_0}{d \theta}$ is presented in Fig. (\ref{Fig_horseshoe_field1}) (black curve). The first guess for the field $f_1$ contains only some small residual order 2 terms. The fist order terms for $f_1$ are canceled by the fitting procedure. This first guess for $f_1$ cannot be compared to the further refined estimation of this field which by construction contains conjugated first order terms from the
$\frac{d f_0}{d \theta}$ field. As a consequence the first guess for $f_1$ is not presented in Fig. (\ref{Fig_horseshoe_field1}). Note that for the  reconstruction of the initial guess the effect of the PSF has been neglected. It is possible to apply a PSF correction for the estimation of the first guess. However at this level the effect of this correction is rather small compared to other uncertainties like for instance the non-circular source terms, or the higher order terms in the modelisation of the lens. As a consequence for the construction of the guess solution the PSF correction was not applied.
\subsection{Refining the solution and converging to the minimal solution.}
\label{Sec_refine_h}
\begin{figure*}
\includegraphics[width=18cm,angle=0]{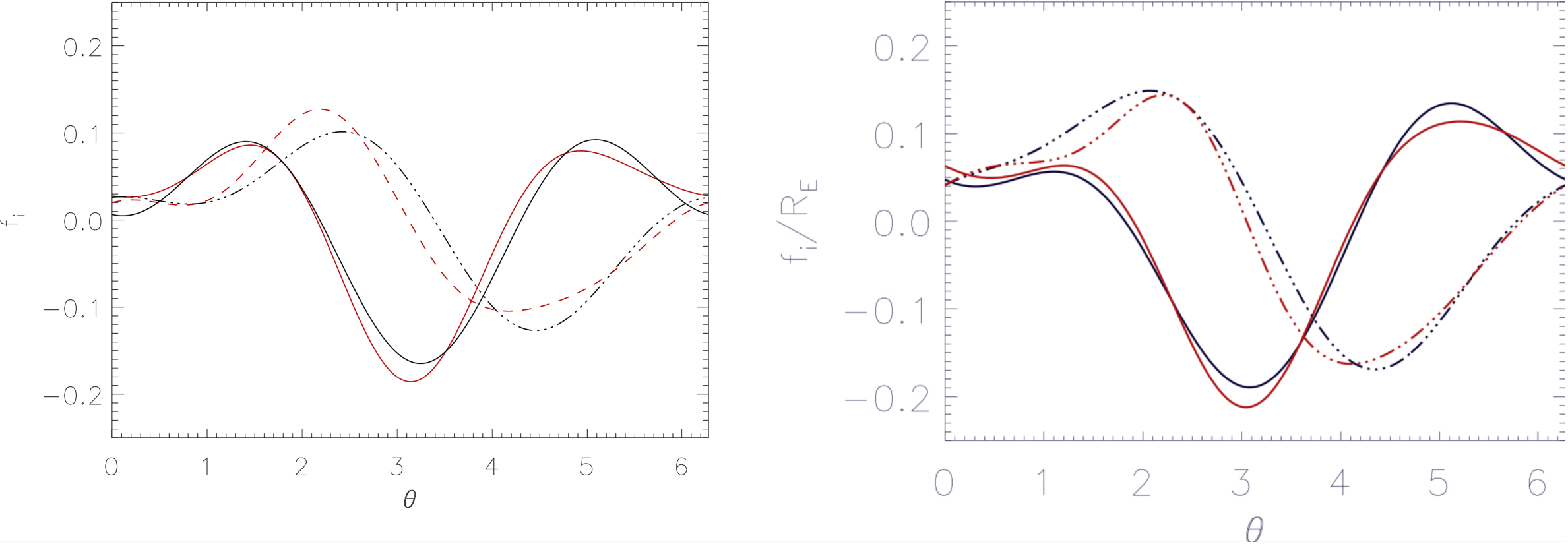}
\caption{A comparison of the current automated solution (left side) to the 2017 solution.
 (right side). The field solution at order 2 (blue color) and at order 3 (red color).  The field $\frac{d f_0}{d \theta}$ is represented by solid lines while $f_1$ is represented by dotted lines.}
\label{Fig_field_sol}
\end{figure*}

\begin{figure*}
\includegraphics[width=18cm,angle=0]{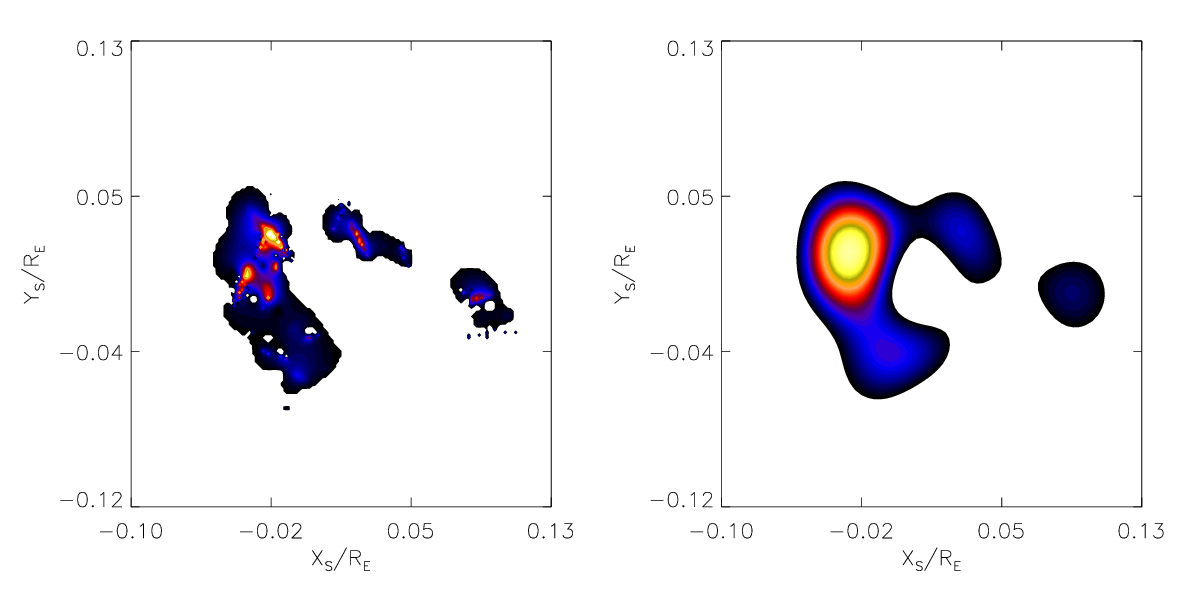}
\caption{The source solution corresponding to the minimal solution obtained in Sec. (\ref{Sec_refine_h}). The basic source reconstruction obtained by fitting the Gaussian polynomial expansion at order 8 (right side) in presented next to the refined source construction (see Sec. \ref{Sec_source}).}
\label{Fig_source0}
\end{figure*}

At this level a full reconstruction of the image with a complex non-circular lens model is performed. The PSF convolution of the source image is also performed by using the empirical model derived by \cite{Anderson}. The fitting procedure is the same as the procedure already used for the simulations. The parameters of the fit are described in Sec. (\ref{subsec_params}), the source reconstruction procedure is detailed in Sec. (\ref{subsec_source}) and the final $\chi^2$ estimation is presented in Sec. (\ref{subsec_chi2}). The source model use an expansion to order 8 in 2D Gaussian polynomials functions (see Eq. (\ref{Eq_source})). The non-linear minimization procedure starts with the initial guess and converge to the best fit at order 2. Then starting applying the same minimization procedure but starting this time from the best solution at order 2, the best solution at order 3 is obtained. Iterating the procedure and computing the best solution at order 4 does not produce any significant improvement of the $\chi^2$. As a consequence the minimal solution corresponds to order 3 (see \cite{Alard2023}). An important basic principle of the automated reconstruction presented here is that the procedure should apply to large scale survey like Euclide. Consequently the procedure has to be simple fast reliable and accurate. It means that the method should converge towards the solution by using source models with a limited degree of complexity. The complexity should be sufficient to allow some degree of flexibility in the modeling. Increasing the complexity of the source may slightly improve the $\chi^2$ of the fit but cause very little change in the modeling of the lens. This is indeed the case for this particular lens, going beyond order 8 for the source model does not change the lens model. Even going from a source model at order 4 does not produce much change in the final result for the lens model. As a consequence we will stop the modeling of the lens here and will not try to refine the lens model by exploring more complicated source model. The goal of this approach is to provide the simplest possible and most efficient in terms of computing method. The overall computing time on a basic PC computer for this lens is only of about 10 minutes for the whole process which make it suitable for large scale search in EUCLIDE like surveys.
The solution for the fields is presented in Fig. (\ref{Fig_field_sol}), while the solution for the source is presented in Fig. (\ref{Fig_source0}). A comparison with the former solution from \cite{Alard2017} (see Fig. (\ref{Fig_field_sol})) indicates that the solutions are close, the main difference between the 2 solutions are due to a difference in the centering of the main circular un-perturbed potential. The order of the difference in centering is of about 1 percent of the Einstein radius in the X coordinate and of about 2 percent in the Y direction. Note that in  this new solution the center of the un-perturbed potential has been refined during the fitting procedure while it was kept constant in the former solution. Actually in the former solution the center of the un-perturbed potential was assumed to be identical to the center of elliptical galaxy inside the arc.
%
%
%
%
\begin{figure*}
\includegraphics[width=18cm,angle=0]{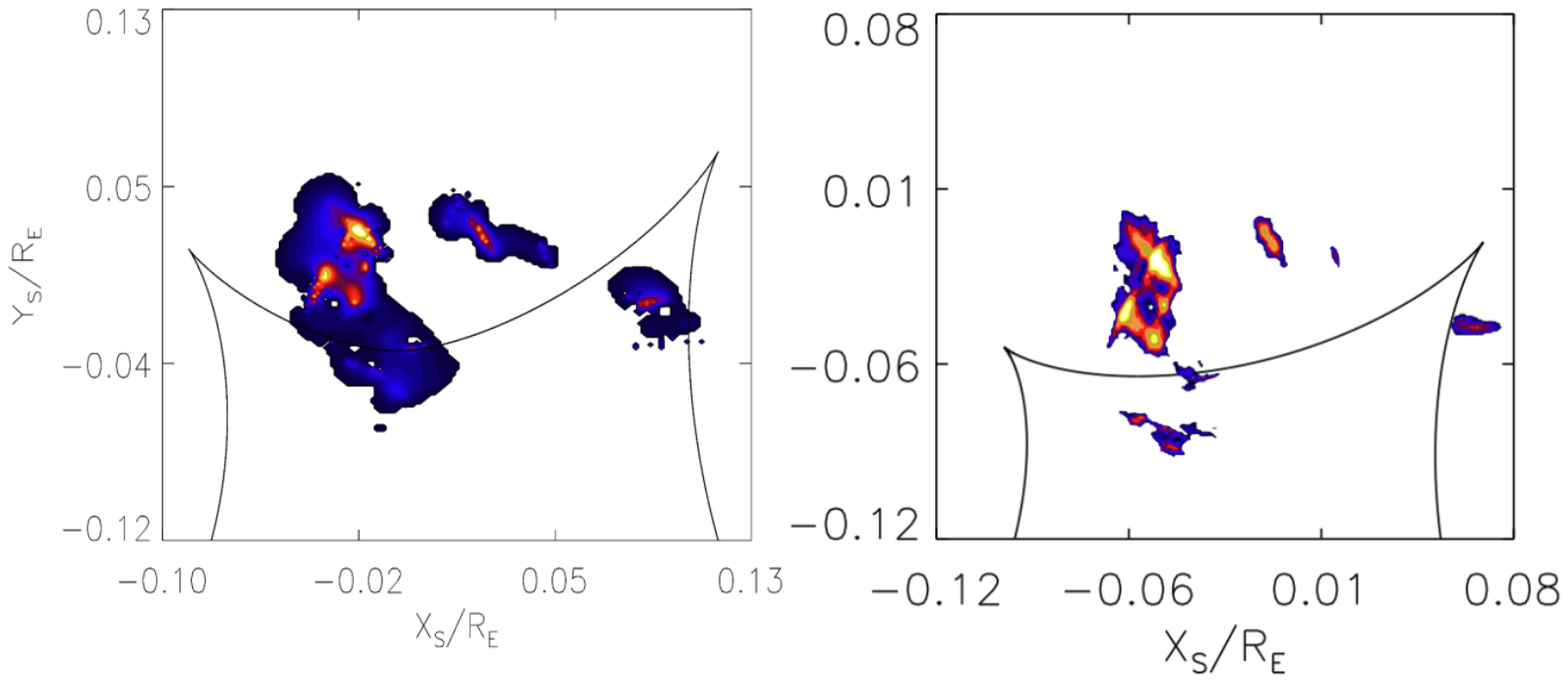}
\caption{The solution in the source plane for the current automated reconstruction (left) next to the 2017 solution. 
The caustics are slightly larger in the current solution but this is due mostly to a a different normalization. In effect the Einstein radius for the current solution is somewhat smaller. Note that in the already published solution 
 the Einstein radius was a fixed quantity, while here its value is refined and optimized in the fitting procedure. }
\label{Fig_caustics_sol}
\end{figure*}
\subsection{A method for an accurate source reconstruction given an optimal lens solution.}
\label{Sec_source}
\begin{figure*}
\includegraphics[width=14cm,angle=0]{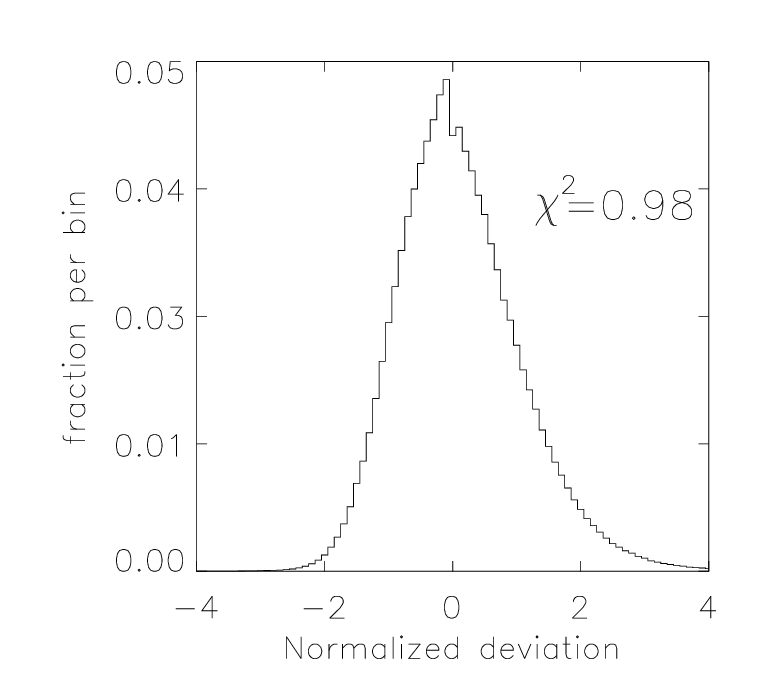}
\caption{The histogram of the residual image normalized by the noise. The residual is obtained by subtracting the data with the model obtained in Sec. (\ref{Sec_source}).}
\label{Fig_noise_hist}
\end{figure*}

\begin{figure*}
\includegraphics[width=16cm,angle=0]{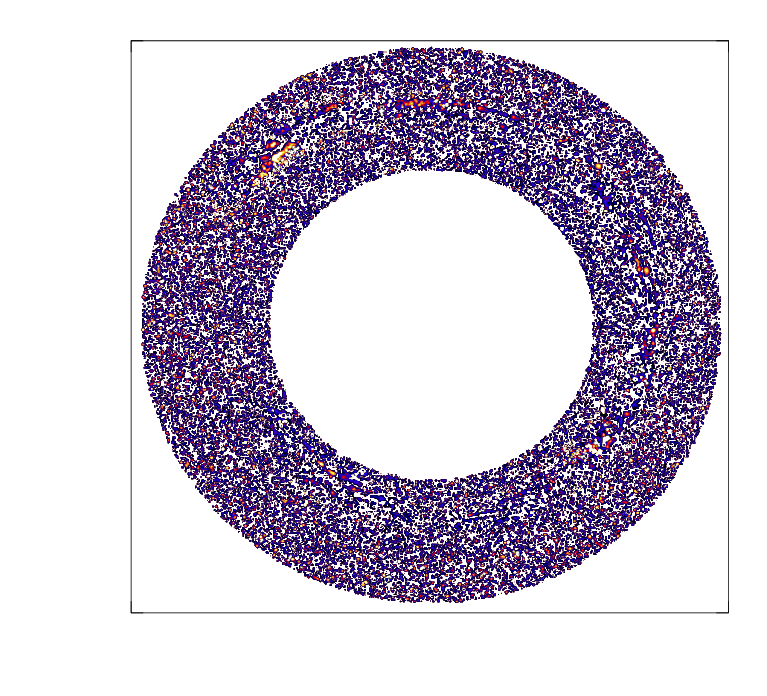}
\caption{The residual image normalized by the noise. The residual is obtained by subtracting the data with the model obtained in Sec. (\ref{Sec_source}).}
\label{Fig_noise_dev}
\end{figure*}

Starting for the minimal solution at order 3 for the lens obtained in Sec. (\ref{Sec_refine_h}) we will now present a procedure to reconstruct the source at high accuracy on a fine grid. In this procedure the lens model is not refined and kept fixed. The source reconstruction method is non-parametric and estimate directly the source model by projecting an image of the arc in the source plane. The problem is to take into account the effect of the image convolution with the PSF. The correction for the PSF convolution is performed iteratively using the Van Cittert method. In practice this means that the first guess will be obtained directly by transforming the data from the image plane to the source plane. Once this first source estimate has been obtained the corresponding image is computed and convolved with the PSF. An estimate of the correction is the difference between the convolved source image and the original data. This correction is then transported back to the source plane and added to the source. The process is iterated until convergence, with a typical number of 10 iterations. It is important to note that this deconvolution process tends to amplify noise and that a regularization process is needed. The regularization is obtained by filtering the source image at each step using a wavelet method. The wavelet method operates by decomposing the image in a number of planes with increasing scales. If the noise is Gaussian the wavelet filtering operates by cutting amplitudes below a certain threshold in each plane. However if the noise is not Gaussian the image must be transformed so that the noise is Gaussian. For Poissonian noise this is a simple operation, and this is the model that was used here. Once the filtering has been performed by cutting amplitude below a $3 \sigma$ threshold in each plane, the transformation is inverted and the plane are summed to reconstruct the wavelet filtered image. This wavelet filtering procedure guaranties the stability of the deconvolution process and avoid an un-necessary and un-controlled amplification of the noise. The solution obtained by this procedure is presented in Fig. (\ref{Fig_horseshoe_images}) for the image, and in Fig. (\ref{Fig_caustics_sol}) for the source. The comparison presented in Fig. (\ref{Fig_caustics_sol}) indicates that the new solution and the former solution from \cite{Alard2017} are very close except for difference in the normalization radius (Einstein radius) and minor differences in the centering of the main circular un-perturbed potential. The noise in the data after the subtraction of the model obtained here is very close to the statistical expectation (see Fig. (\ref{Fig_noise_hist})). The $\chi^2$ in the residual image is 0.98 for the whole fitting area, while it is slightly higher when an area with high values in the original data is selected. For instance by selecting pixels with values higher than half of the maximum pixel value in the arc the $\chi^2$ rise to 1.15. The residual image normalized by the noise expectation is presented in Fig. (\ref{Fig_noise_dev}). It is apparent in this image that as already noted there are some slightly larger residuals in the areas where the arc image has its higher values. This is due to the fact that the actual source has very sharp details at a very small scale and that even the higher resolution reconstruction of the source cannot represent these features. Going to even higher resolution for the source reconstruction will face the problem of the degeneracy in the reconstruction. Only a fine tuned specific model would solve the issue but this is not the purpose of this automated reconstruction.
%
%
\section{Conclusion.}
In this paper systems of gravitational arcs of increasing complexity were analyzed and successfully reconstructed. Even in the case of a fourth order lens potential and a strong external shear an accurate reconstruction of the potential and source was achieved. The same complex highly non symmetrical source model was used for all the simulations. 
This analysis demonstrates that the method developed in this paper has the ability to perform automated reconstructions of gravitational arcs systems. The reconstructions are direct due to the nature of the singular perturbative formalism and as a consequence insensitive to degeneracy issues. Considering the rapidly increasing number of good quality images of gravitational arcs systems this package of programs is of particular interest. The method offers the the possibility to reconstruct quickly a large number of arcs using a common universal model. The universality of the potential modeling allows the reconstruction of statistical quantities for a large number of arcs. This possibility is of particular interest 
when the data from incoming high quality sky surveys like Euclide will be available. 
\section*{Data Availability}
No data sets were generated or analyzed during the current study.
\section*{Acknowledgements}
The author would like to thank the referee for making interesting suggestions.
\end{document}